\documentclass[letter,scriptaddress,twocolumn, prl,showkeys]{revtex4}
\usepackage{graphicx}

\def\beq{\begin{equation}}
\def\eeq{\end{equation}}
\def\beqar{\begin{eqnarray}}
\def\eeqar{\end{eqnarray}}

\def\para{\parallel}

\newcommand{\diff}[2]{\frac{d#1}{d#2}}

\newcommand{\pdiff}[2]{\frac{\partial#1}{\partial#2}}

\newcommand{\pdt}{\partial_t}

\def\grad{\nabla}

\newcommand{\gradpar}{\grad_\parallel}
\newcommand{\gradperp}{\grad_\perp}

\newcommand{\vpe}{v_{\parallel e}}

\newcommand{\nue}{\nu_{e}}

\newcommand{\nuin}{\nu_{in}}
\newcommand{\kpe}{\kappa_{\parallel e}}

\newcommand{\fmie}{\frac{m_i}{m_e}}

\begin{document}

\title{A Linear Technique to Understand Non-Normal Turbulence Applied to a Magnetized Plasma}

\author{B. Friedman}
\email{friedman@physics.ucla.edu}

\author{T.A. Carter}

\affiliation{Department of Physics and Astronomy, University of California, Los Angeles, California 90095-1547, USA}

\begin{abstract}
In nonlinear dynamical systems with highly nonorthogonal linear eigenvectors, linear non-modal analysis is more useful than normal mode analysis in predicting turbulent properties. 
However, the non-trivial time evolution of non-modal structures makes quantitative understanding and prediction difficult. 
We present a technique to overcome this difficulty by modeling the effect that the advective nonlinearities have on spatial turbulent structures. 
The nonlinearities are taken as a periodic randomizing force with
time scale consistent with critical balance arguments. We apply this technique to a model of drift wave turbulence in the Large Plasma Device (LAPD) 
[W. Gekelman \emph{et al.}, Rev. Sci. Inst. {\bf 62}, 2875 (1991)], where non-modal effects dominate the turbulence.
We compare the resulting growth rate spectra to that obtained from a nonlinear simulation, showing good qualitative agreement, especially in comparison to the eigenmode growth rate spectra.
\end{abstract}

\maketitle

Normal mode analysis -- the calculation of eigenvalues and eigenvectors of a linearized dynamical system -- has been used to solve many problems over the years.
Despite its wide-ranging success, it has failed in important instances, particularly in predicting the onset of subcritical turbulence in hydrodynamic flows. 
The reason for this failure was explained in the early 1990's when Trefethen and others attributed the pitfalls of normal mode analysis to the non-normality of linear operators of
dynamical systems~\cite{trefethen1993,schmid2007}. A non-normal operator has 
eigenvectors that are not orthogonal to one another. One consequence of eigenvector nonorthogonality is that even when all eigenvectors decay exponentially under linear evolution, 
superpositions of eigenvectors can grow, albeit transiently.
In other words, certain fluctuations of the laminar state can access free energy from background gradients even though normal mode fluctuations cannot.
When combined with nonlinear effects, this allows for sustained subcritical turbulence.
Such behavior is obscured by traditional normal mode analysis, which only effectively describes the long time asymptotic behavior of fluctuations under  
action of the linear operator. Transient growth, which can dominate turbulent evolution, can be discovered only through non-modal calculations.

Non-modal analysis has been embraced by the hydrodynamics community over the past two decades in the attempt to explain and predict subcritical turbulence. But the plasma community
generally relies on normal mode analysis to inform turbulent predictions and observations, with a few exceptions~\cite{camargo1998,camporeale2010,schekochihin2012}. 
Furthermore, non-modal treatments have generally been explanatory rather than predictive and have centered around the transition to turbulence in subcritical systems rather 
than on properties of fully-developed turbulence.
This paper takes up the task of developing an approach to understand highly non-normal (implying highly collisional~\cite{camargo1998}) 
turbulent properties using only non-modal linear calculations with the goal of 
ultimately making quantitative predictions.  Our approach is to calculate
an average growth rate of turbulent fluctuations due to linear
processes, specifically due to transient growth.  We model the
turbulent steady state as a series of processes:  (1) the turbulence starts as a spatially random state, (2) linear transient growth deterministically amplifies the turbulent energy (or
decrease it in wavenumber ranges where linear damping dominates), (3) nonlinear transfer sets in at a specified timescale, terminating the
transient growth process and re-randomizing the turbulent state (at which point the cycle repeats).    Optimally, the timescale for the final step would be the
nonlinear decorrelation time of the turbulent system, but in order to enable predictive capability, we employ critical balance
arguments to use a characteristic \emph{linear} time. Since there is no obvious single linear time scale, we test several and compare the results to determine which works best.
The procedure ultimately produces growth rate spectra that can be used
to predict turbulent properties such as saturation levels and transport rates through mixing length arguments.
While the concepts behind this technique are general enough to be applied to various nonlinear dynamical systems, the details vary for each case, 
so we restrict our treatment to one particular turbulence model. For
this model, the technique does reasonably well in reproducing the turbulent growth rate spectrum of the direct 
nonlinear simulation, especially in comparison to the linear eigenmode spectrum.

The model we use describes highly collisional pressure-gradient-driven turbulence in the uniformly magnetized, cylindrical plasma
produced by the Large Plasma Device (LAPD)~\cite{gekelman1991}. 
We use a reduced Braginskii 2-fluid model~\cite{Popovich2010a,Popovich2010b,Umansky2011,friedman2012b,friedman2013}:

\beqar
\label{ni_eq}
\pdt N = - {\mathbf v_E} \cdot \grad N_0 - N_0 \gradpar \vpe + S_N + \{\phi,N\}, \\
\label{ve_eq}
\pdt \vpe = - \fmie \frac{T_{e0}}{N_0} \gradpar N - 1.71 \fmie \gradpar T_e  \nonumber \\
+ \fmie \gradpar \phi - \nue \vpe + \{\phi,\vpe \}, \\
\label{rho_eq}
\pdt \varpi = - N_0 \gradpar \vpe  - \nuin \varpi + \{\phi,\varpi \}, \\
\label{te_eq}
\pdt T_e = - {\mathbf v_E} \cdot \grad T_{e0} - 1.71 \frac{2}{3} T_{e0} \gradpar \vpe \nonumber \\
+ \frac{2}{3 N_0} \kpe \gradpar^2 T_e  - \frac{2 m_e}{m_i} \nue T_e  +  S_T + \{\phi,T_e\},
\eeqar
where $N$ is the density, $\vpe$ the parallel electron velocity, $\varpi \equiv \gradperp \cdot (N_0 \gradperp \phi)$ the potential vorticity,
$T_e$ the electron temperature, ${\mathbf v_E}$ the ${\mathbf E} \times {\mathbf B}$ velocity, and $S_N$ and $S_T$ density and temperature sources. All lengths are
normalized to the ion sound gyroradius $\rho_s$, times to the ion cyclotron time $\omega_{ci}^{-1}$, velocities to the sound speed $c_s$, 
densities to the equilibrium peak density, and electron temperatures and potentials to the equilibrium peak electron temperature. 
The profiles $N_0$ and $T_{e0}$ and other parameters are taken from experimental measurements~\cite{schaffner2012,friedman2012b,friedman2013}. 

The equations are global and we retain advective nonlinearities, which are written with Poisson brackets, but we neglect other nonlinear terms. 
We add artificial diffusion and viscosity terms with small numerical coefficients ($10^{-3}$)
to ensure numerical stability in nonlinear simulations, which are performed with the BOUT++ code~\cite{dudson2009}. We use periodic axial and zero value radial 
boundary conditions. Further details of the model, including validation studies, may be found in the references~\cite{Popovich2010a,Popovich2010b,Umansky2011,friedman2012b,friedman2013},
though we mention here that the model does very well in reproducing the statistical properties of the experimentally-observed turbulence.

The nonlinear simulation reveals a fascinating property of the turbulence -- 
it is dominated by a nonlinear instability process despite being linearly unstable to drift waves~\cite{friedman2012b,friedman2013}.
The nonlinear instability, which was discovered by Drake et al.~\cite{drake1995}, works as follows: 
magnetic-field-aligned ($k_\para=0$) convective filaments transport density across the equilibrium density gradient, setting up $k_\para=0$ density filaments. 
These filaments are unstable to secondary drift waves, which grow on the periphery of the filaments. 
These drift waves, which have finite $k_\para$, nonlinearly couple to one another and generate new convective filaments.

Although the instability is called a nonlinear instability, the first part of the mechanism -- the transport of background density by the convective filaments -- is a linear one.
In fact, the other parts of the mechanism are driven by energetically conservative nonlinear interactions, 
meaning that the convective transport is the only step responsible for energy injection into the fluctuations.

Deriving an equation for the evolution of the energy from Eqs.~\ref{ni_eq}-\ref{te_eq}~\cite{friedman2012b,friedman2013}, we may symbolically write

\beq
\label{dEdt_def}
\diff{E(m,n)}{t} = \diff{E_{l}(m,n)}{t} + \diff{E_{nl}(m,n)}{t}
\eeq
where $m$ and $n$ represent the azimuthal and axial Fourier mode numbers. 
$\diff{E_{l}(m,n)}{t}$ comes from the linear terms in Eqs.~\ref{ni_eq}-\ref{te_eq}. $\diff{E_{nl}(m,n)}{t}$ comes from the nonlinear terms. 
$\diff{E_{l}(m,n)}{t}$ represents the injection (or dissipation) of energy into the fluctuations from the free energy in the equilibrium gradients.
$\diff{E_{nl}(m,n)}{t}$ accounts for the energy exchange between fluctuations with different $m,n$ and it is conservative: $\sum_{m,n} \diff{E_{nl}(m,n)}{t} = 0$.
Moreover, in quasi-steady state turbulence, the rate of energy injection (or dissipation) into the fluctuations at each $m,n$ by the linear terms must be balanced by
the rate of energy removal (or deposition) from the nonlinear terms:

\beqar
\label{steady_state}
\gamma(m,n) \equiv  \displaystyle\lim_{T \to \infty} \frac{1}{T} \int_0^T \frac{dE(m,n)/dt}{2 E(m,n)} dt \nonumber \\
= \displaystyle\lim_{T \to \infty} \frac{{\rm{Log}}\left[ E(T)/E(0) \right]}{T} = 0 \quad {\rm{in \ steady \ state}}.
\eeqar

\begin{figure}
\centerline{\includegraphics[width=0.55\textwidth]{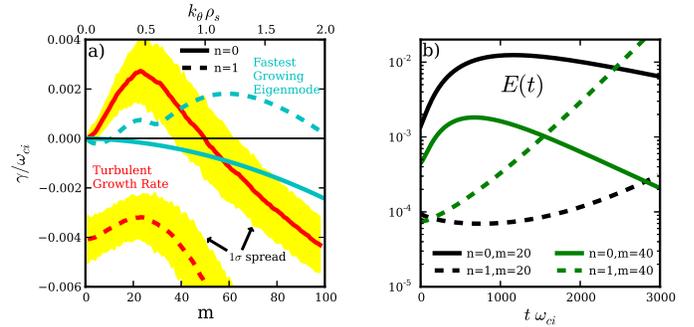}}
\caption{
a) Linear and turbulent growth rate spectra for $n=0$ (solid lines) and $n=1$ (dashed lines) Fourier components. The linear growth rates are those of the least stable eigenmodes,
while the turbulent growth rates represent $\pdiff{E_l}{t}/2 E$ from the nonlinear simulation. The shaded region marks the $1 \sigma$ spread in the turbulent spectrum,
obtained from the distribution of growth rates in the nonlinear simulation.
b) Linear evolution of energy starting from a turbulent initial state. The $n=0$ curves have an initial period of transient growth before exponentially decaying.}
\label{n0_n1_growth}
\end{figure}

From, Eqs.~\ref{dEdt_def} and~\ref{steady_state}, it follows that $ \gamma(m,n) = \gamma_l(m,n) + \gamma_{nl}(m,n) = 0$.
The rate of energy injection $\gamma_l(m,n)$ from the equilibrium gradient into the fluctuations is a quantity of great interest~\cite{friedman2012b,terry2006b}. 
When positive for some wavenumber, turbulence can be
sustained. Additionally, it may be used in a mixing length argument
to predict the turbulent saturation level, 
which can in turn be related to the rate of cross-field transport~\cite{terry2006b}. 
$\gamma_l(m,n)$ may be calculated from the spatial structures of the plasma state variables, so it is always well-defined, even in a turbulent plasma. We plot
$\gamma_l(m,n)$ in Fig.~\ref{n0_n1_growth} a) for both the linear and steady state turbulent stages of our nonlinear simulation. The linear stage, which occurs when fluctuations
are small and exponentially growing has a time-independent $\gamma_l(m,n)$ equal to $\gamma_s(m,n)$ -- the ``spectral'' growth rate of the fastest growing eigenmode at each $m,n$.
During the turbulent stage, $\gamma_l(m,n)$ is time-dependent (indicated by the $1 \sigma$ spread), and generally much different than $\gamma_s(m,n)$. Significantly,
the turbulent $\gamma_l$ is positive at $n=0$ and low $m$ despite the fact that all linear eigenmodes have $\gamma_s < 0$ for $n=0$. This is a manifestation
of non-normality, for in normal systems, $\gamma_l(m,n) \le \gamma_s(m,n)$. Physically, it is the manifestation of the nonlinear
instability, specifically the linear part of the mechanism in which convective filaments drive density filaments from the equilibrium density gradient.

This convective transport of density filaments is akin to the paradigmatic ``lift-up'' mechanism in hydrodynamic shear flows whereby streamwise vortices drive streamwise 
streaks~\cite{trefethen1993,krommes1999}.
Both are transient growth processes. We see this in our simulations by following the evolution of the energy of several $m,n$ modes after turning off the nonlinearities in an already
turbulent simulation. A few representative modes are shown in Fig.~\ref{n0_n1_growth} b). The linear transient growth of the filamentary $n=0$ structures is evident as their modes 
grow transiently before decaying exponentially at the rate of their least stable eigenmode. Such behavior is indicative of non-modal behavior. It must be since 
all $n=0$ linear eigenmodes are stable. Notice also that the $n=1,m=20$ mode decays transiently before growing exponentially with growth rate of the most unstable eigenmode. This transient
decay is also a non-modal result since $\gamma_s(n=1,m=20) > 0$.

Since transient growth is a purely linear phenomenon, it has been a goal of researchers to understand and predict the onset of subcritical turbulence using only linear, non-modal calculations. 
In our system, the turbulence is not subcritical in the traditional sense, but it has a subcritical component because $\gamma_l(m<50,n=0) > 0$, yet $\gamma_s(m,n=0) < 0$. 
It is our goal, then, to use linear non-modal calculations to understand this behavior and to move toward predictive capability of $\gamma_l$.
To accomplish this, we make the ansatz that nonlinearities randomize the turbulent spatial structure at each wavenumber on a time scale of 
one eddy decorrelation time, while the linearities evolve the spatial structures deterministically. From this we can calculate a non-modal $\gamma_{\rm{n-m}}$ spectrum
which is our prediction of the turbulent $\gamma_l$ spectrum. To illustrate,
we begin by taking Eqs.~\ref{ni_eq}-\ref{te_eq} and Fourier decomposing in the azimuthal and axial directions.  Then, we discretize in the radial direction and
approximate radial derivatives with finite differences.  The resulting system of equations may be written in matrix form:

\beqar
\label{mat_eq}
\mathbf{B}_{m,n} \diff{\mathbf{v}_{m,n}(t)}{t} = \mathbf{C}_{m,n} \mathbf{v}_{m,n}(t) \nonumber \\
- \sum_{m',n'}  \mathbf{v}_{E,m-m',n-n'} \cdot \gradperp \left( \mathbf{B}_{m',n'} \mathbf{v}_{m',n'}(t) \right),
\eeqar
where $\mathbf{v}_{m,n} = \left( N(r), \vpe(r), \phi(r), T_e(r) \right)_{m,n}^{T}$ is the state of the system,
and $\mathbf{B}_{m,n}$ and $\mathbf{C}_{m,n}$ are coefficient matrices that include the equilibrium information and finite difference coefficients. The first term on the RHS represents the linearities
and the second term the nonlinearities. Note that for each $m,n$, there exist $4 N_r$ linearly independent, but nonorthogonal eigenvectors, where $N_r$ is the number of radial grid points. 
Hence forth, we drop the $m,n$ Fourier subscripts.

In order to use non-modal analysis to calculate growth rates and other measures, one must choose a norm and inner product with which to work. While any choice of
inner product is possible, a physically relevant one such as an energy inner product is generally preferred~\cite{camargo1998,schmid2007,camporeale2010}. 
Recall that the inner product of two vectors may be written $\left< \mathbf{x},\mathbf{y} \right> = \mathbf{y}^{\dagger} \mathbf{M} \mathbf{x}$.
We choose $\mathbf{M}$ so that $||\mathbf{v}||^2 = \left< \mathbf{v},\mathbf{v} \right> = E$. Furthermore, it is convenient in
computations to use the $L_2$-norm, $||\mathbf{u}||_2^2 = \sum_i |u_i|^2$. This can be accomplished through the change of variables $\mathbf{u} = \mathbf{M}^{\frac{1}{2}} \mathbf{v}$.
Then the linear portion of Eq.~\ref{mat_eq} becomes

\beq
\label{lin_eq_A}
\diff{\mathbf{u}}{t} = \mathbf{A} \mathbf{u},  \quad \rm{where} \ \mathbf{A} = \mathbf{M}^{\frac{1}{2}} \mathbf{B}^{-1} \mathbf{C} \mathbf{M}^{-\frac{1}{2}}.
\eeq
The solution of Eq.~\ref{lin_eq_A} is $\mathbf{u}(t) = e^{\mathbf{A} t} \mathbf{u}(0)$, which
depends on the initial condition $\mathbf{u}(0)$. For purposes of turbulent growth rate prediction, we are interested in
the behavior of $G(t) = E(t)/E(0) = ||\mathbf{u}(t)||^2/||\mathbf{u}(0)||^2$.


It is common practice in normal mode analysis to look for the least stable eigenmode. 
For the non-modal case, it is common to study the properties of $G_{\rm{max}}(t) = ||e^{\mathbf{A} t}||$ because if this is greater than unity at any time, fluctuations may be amplified, 
leading to subcritical turbulence~\cite{trefethen2005,schmid2007}.
However, it can be misleading to study only $G_{\rm{max}}(t)$ when predicting specific properties of turbulence because
$G_{\rm{max}}(t)$ is only the upper envelope of all possible $G(t)$ curves. No one particular initial condition $\mathbf{u}(0)$ evolves along $G_{\rm{max}}(t)$. 
Furthermore, it isn't obvious what kind of spatial structures will come to dominate a turbulent system.
In non-normal systems, unlike in normal systems, optimal structures don't amplify themselves, rather, they evolve while increasing the total fluctuating energy.

\begin{figure}
\centerline{\includegraphics[width=0.45\textwidth]{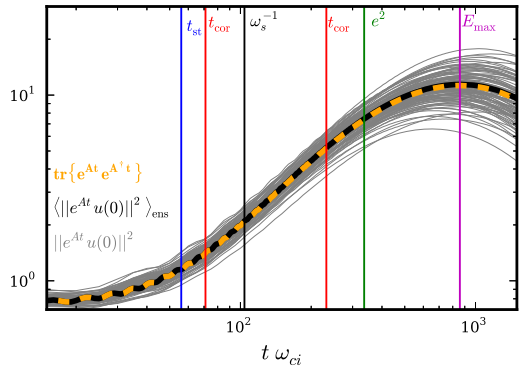}}
\caption{An ensemble of initially randomized growth ratio curves (solid gray lines, with the solid black line their average, and the dashed orange line the average as calculated from Eq.~\ref{E_t_ensemble_avg}).
The vertical lines indicate various linear time scales.}
\label{m20n0_E_ev}
\end{figure}

We contend that the key to understanding and predicting turbulent properties through non-modal analysis is 
to successfully model the effect that the nonlinearities have on the transient linear processes. 
To this effect, we note that the advective nonlinearity in Eq.~\ref{mat_eq} has the form of the state vector divided by a time $\tau_{nl} \sim (v_E k_\perp)^{-1}$. This nonlinear
time scale is generally associated with the eddy turnover or decorrelation time. We therefore present a heuristic model of the nonlinearities 
as a randomizing force that acts on this characteristic nonlinear time scale. Again, this model is one in which 1) the turbulence begins as a random state, 
2) evolves linearly for a time $\tau_{nl}$, and 3) randomizes by nonlinear energy transfer, at which point the steps repeat.
In practice, we implement this model by starting with an ensemble of random initial conditions, which we evolve linearly for a time $\tau_{nl}$, 
and then take the time and ensemble averaged growth rate of these curves.
This procedure does not require actual linear simulations from an ensemble of random initial conditions. To see this, recall that the time evolution of the energy from an initial condition is

\beq
\label{E_t_from_u0}
E(t) = ||e^{\mathbf{A} t} \mathbf{u}(0)||^2 = e^{\mathbf{A} t} \mathbf{u}(0) \mathbf{u}^{\dagger}(0) e^{\mathbf{A}^{\dagger}t}.
\eeq
If the $\mathbf{u}(0)$ in the ensemble are random with uncorrelated components and normalized to unity, it follows that~\cite{camargo1998}

\beq
\label{E_t_ensemble_avg}
\left< E(t)/E(0) \right>_{{\rm{ens}}} = \frac{1}{4 N_r} {\rm{tr}} \{ e^{\mathbf{A} t} e^{\mathbf{A}^{\dagger}t} \}.
\eeq
We show the validity of this statistical averaging by plotting an
ensemble of $1000$ curves generated with different random initial conditions in Fig.~\ref{m20n0_E_ev} along with their expected average from Eq.~\ref{E_t_ensemble_avg} and their actual average,
which agree well.
One point to note is that in global equation sets like ours -- where we discretize and use finite differences in the radial direction rather than a Fourier decomposition -- randomizing $\mathbf{u}(0)$
amounts to setting the initial $k_r$ spectrum to a step function that goes to zero at the Nyquist wavenumber. This means that $\left< E(t)/E(0) \right>_{{\rm{ens}}}$ may depend on $N_r$, so we must be careful in
choosing $N_r$ for this analysis. We choose $N_r$ such that the grid spacing equals $\rho_s$, the general Nyquist wavenumber of drift wave simulations~\cite{scott1992}.

Mathematically, our procedure is to calculate $\gamma_{\rm{n-m}}$ by the following formula:

\beq
\label{gamma_l_calc}
\gamma_{\rm{n-m}} = \frac{1}{\tau_{nl}} \int_0^{\tau_{nl}} \frac{\pdiff{E(t)}{t}}{2 E(t)} dt = \frac{1}{2 \tau_{nl}} {\rm{Log}} \left[ \frac{E(\tau_{nl})}{E(0)}\right]
\eeq
where $E(t)$ is the ensemble averaged energy calculated from Eq.~\ref{E_t_ensemble_avg}, and $E(0) = 1$ by our normalization.
In order to move toward predictive capability, we must estimate $\tau_{nl}$ with only knowledge of linear (modal or non-modal) information. 
We thus invoke the conjecture of \emph{critical balance}, which posits that the nonlinear time scale equals the linear time scale at all spatial scales~\cite{schekochihin2012}. This follows from 
the previously derived steady-state result: $\gamma_l = - \gamma_{nl}$. 

Now there are several linear time scales that we may choose to test. We label these times in Fig.~\ref{m20n0_E_ev}
for the case of $n=0, m=20$. The first linear time is the linear eigenmode frequency, labeled
$\omega_s^{-1}$. However, $\omega_s = 0$ for $n=0$ linear eigenmodes, so we are forced to use $\omega_s$ for the fastest growing $n=1$ eigenmode at each $m$ to get a meaningful time scale.
Second is the parallel free-streaming time of the electrons $t_{\rm{st}} = L_\para/v_{t e}$ often cited in critical balance arguments~\cite{barnes2011}. Third is the time before modal effects
take over, which can be approximated as the time when $E(t)$ turns over. As seen in Fig.~\ref{n0_n1_growth} b), $E(t)$ can be either a maximum or minimum before turning over. 
We label this time $E_{\rm{max}}$. 
Fourth, we use the steady-state condition $\gamma_l = - \gamma_{nl}$ and the approximation $\tau_{nl} \sim 1/|\gamma_{nl}|$ to get $\tau_{nl} = 1/|\gamma_{\rm{n-m}}|$. 
Inserting this into Eq.~\ref{gamma_l_calc} gives

\beq
\label{fourth_gamma_cond}
 {\rm{Log}} \left[ E(1/\gamma_{\rm{n-m}})\right] = \pm 2 \ \rightarrow E(1/\gamma_{\rm{n-m}}) = e^{\pm 2}.
\eeq
In other words, we find the time at which $E(t)$ grows to the value of $e^2$ or decays to the value of $e^{-2}$, and then use this time to get $\gamma_{\rm{n-m}}$. We label this time $e^2$. 
In Fig.~\ref{m20n0_E_ev},
we also indicate two times labeled $t_{\rm{cor}}$, which when inserted into Eq.~\ref{gamma_l_calc} give the ``correct'' $\gamma_l(m=20,n=0)$ calculated directly from the nonlinear simulation.

\begin{figure}
\centerline{\includegraphics[width=0.55\textwidth]{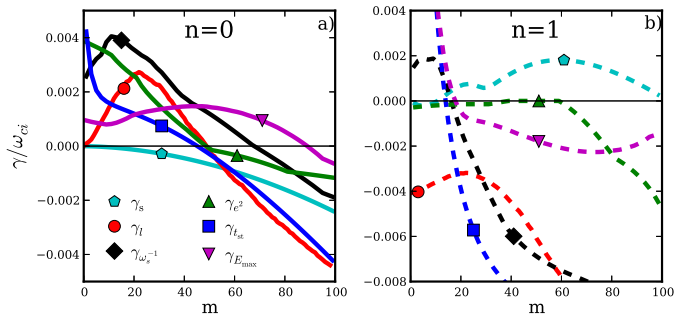}}
\caption{The growth rate spectra of the spectral growth rate $\gamma_s$, turbulent simulation growth rate $\gamma_l$, and growth rates calculated from the non-modal procedure in Eq.~\ref{gamma_l_calc}.
The growth rates as a function of $m$ for a) $n=0$ and b) $n=1$.}
\label{gamma_comparisons}
\end{figure}

In Fig.~\ref{gamma_comparisons}, we compare the $\gamma_{l}$ spectrum from the nonlinear simulation to $\gamma_{n-m}$ from the non-modal procedure using the four different linear time scales. We also show
$\gamma_s$ for reference. All of the non-modal growth rates are positive at $n=0$ for low $m$, like $\gamma_l$ from the simulation and unlike $\gamma_s$, indicating that the non-modal analysis can reveal
what normal mode analysis cannot in this turbulent system. Furthermore, the choice of linear time scale does not significantly affect the qualitative picture.

On the other hand, the non-modal analysis does not always
predict complete stability at $n=1$ for all choices of linear time scale, but it does indicate that $n=1$ modes can dissipate rather than inject fluctuation energy despite the presence of unstable
linear eigenmodes. Finally, we find that the linear eigenmode time scale $\omega_s^{-1}$ gives the best match to $\gamma_l$ for this model.

In summary, we present a procedure for calculating the turbulent growth rate spectrum using non-modal linear calculations. 
In the case of a simulation of an LAPD experiment, this procedure captures the behavior of a nonlinear instability that dominates the dynamics of the turbulence.  In general,
non-modal analysis is difficult to quantify and make predictive, but using some simple nonlinear modeling, we have shown
that it may be possible. Future studies will attempt to test this procedure on other turbulence models, and see if it can predict critical parameters for subcritical turbulent onset.

This work was supported by the National Science Foundation (Grant PHY-1202007)



\end{document}